\documentclass{elsart}
\usepackage{amsmath,latexsym,graphicx,psfrag,caption2}
\usepackage{dsfont}

\def\HPL{{\tt HPL}}
\begin{document}
\newcommand{\arctanh}{\mathrm{arctanh}}
\newcommand{\arccot}{\mathrm{arccot}}
\newcommand{\arccsc}{\mathrm{arccsc}}
\newcommand{\sgn}{\mathrm{sgn}}
\newcommand{\prim}[1]{{#1^{\prime}}}
\newcommand{\tr}{\mathrm{Tr}}
\newcommand{\naeher}{\!\!\!}
\newcommand{\vect}[1]{\vec #1}
\newcommand{\ddd}[1]{\d\vect{#1} }
\newcounter{im}
\setcounter{im}{0}
\def\exampleboxlength{1.15\textwidth}
\newcommand{\example}{\stepcounter{im}\includegraphics{./HPL2Examples_\arabic{im}.eps}}

\begin{frontmatter}


\title{\vspace{-2cm}\hfill {\small\rm ZU-TH 5/07\\\vspace{-0.3cm} \hfill hep-ph/0703052}\vspace{2cm}\\Extension of \HPL~to complex arguments  }
 \author{Daniel Ma\^{\i}tre}
 \ead{maitreda@physik.unizh.ch}
 \ead[url]{http://krone.physik.unizh.ch/~maitreda/HPL/}
 \address{Institut f\"ur Theoretische Physik\\Universit\"at Z\"urich\\Winterthurerstrasse 190\\CH-8057 Z\"urich
}






\begin{abstract}
In this paper we describe the extension of the Mathematica package HPL to treat  harmonic polylogarithms of complex arguments. The harmonic polylogarithms have been introduced by Remiddi and Vermaseren \cite{Remiddi} and have many applications in high energy particle physics.
\end{abstract}

\begin{keyword}
Harmonic polylogarithms\sep multiple zeta values\sep Mathematica

\PACS 02.10.De\sep 02.30.Gp
\end{keyword}
\end{frontmatter}
{\bf PROGRAM SUMMARY}

\begin{small}
\noindent
{\em Program Title:} HPL                                   \\
{\em Journal Reference:}                                      \\
{\em Catalogue identifier:}                                   \\
{\em Licensing provisions:} none                                   \\
{\em Programming language:} Mathematica                                  \\
{\em Computer:}     All computers running Mathematica                   \\
{\em Operating system:}  Operating systems running Mathematica          \\
{\em Keywords:} Harmonic polylogarithms, multiple zeta values, Mathematica \\
{\em PACS:} 02.30.Gp, 02.10.De                                                 \\
{\em Classification:}  4.7 Computational Methods, Other functions                                       \\
{\em Nature of problem:}\\
Computer algebraic treatment of the harmonic polylogarithms which appear in the evaluation of Feynman diagrams
   \\
{\em Solution method:}\\
Mathematica implementation

\end{small}

\newpage


\hspace{1pc}
{\bf LONG WRITE-UP}
\section{Introduction}
The harmonic polylogarithms (HPL) of Remiddi and Vermaseren introduced in \cite{Remiddi} appear in many applications of high energy physics. They are a generalization of the usual polylogarithms \cite{Lewin} and Nielsen polylogarithms \cite{Nielsen}. Due to their definition through iterated integrations, they are useful for constructing solutions of differential equations, as they appear for example in the computation of Feynman integrals. 

They are found in three-loop deeply inelastic splitting and coefficient functions
\cite{Vermaseren:2005qc,Moch:2004xu,Vogt:2004mw,Moch:2004pa}, in two-loop
massive vertex form factors
\cite{Bonciani:2003te,Bonciani:2003ai,Bonciani:2003hc,Bernreuther:2004ih,Bernreuther:2004th,Bernreuther:2005rw,Mastrolia:2003yz,Degrassi:2005mc}, in
two-loop Bhabha scattering
\cite{Bonciani:2003cj,Bonciani:2004gi,Bonciani:2004qt,Czakon:2004wm,Bonciani:2005im,Penin:2005kf,Bonciani:2006qu}, in multi-loop three-point and four-point functions
\cite{Bern:2005iz,Birthwright:2004kk,Aglietti:2004ki,Heinrich:2004iq,Aglietti:2004nj,Aglietti:2004tq,Smirnov:2003vi,Aglietti:2003yc,Gehrmann:2001ck,Gehrmann:2000zt,Gehrmann:2002zr}, in 2- and 3-loop lepton g-2 \cite{Passera:2004bj,Passera:2006gc}, in the Higgs production and decay \cite{Harlander:2005rq,Bernreuther:2005gw,Anastasiou:2006hc,Aglietti:2006tp}, heavy quark forward-backward asymmetry  \cite{Bernreuther:2006vp,Bonciani:2006eu} and form factors \cite{Bernreuther:2006yt}, large-$x$ limit of parton evolution \cite{Dokshitzer:2005bf} and various loop calculations \cite{Cachazo:2006tj,Anastasiou:2005cb,Bern:2006ew} and in more
formal developments \cite{Blumlein:2005jg}. 

HPLs appear in the expansion of hypergeometric functions, with real arguments for integer parameters \cite{Weinzierl:2002hv,weinzierl1,Weinzierl:2004bn,Huber:2005yg} and imaginary arguments for hypergeometric functions with half-integer parameters \cite{Kalmykov:2006hu,hypexp2}. HPLs of complex arguments also appear in two-loop massive master integrals \cite{Anastasiou:2006hc,Aglietti:2006tp} and in the anomaly contribution to the heavy quark form factors \cite{Bernreuther:2005rw}. HPLs on the unit circle are related to the Clausen function and the generalized log-sine integrals \cite{Lewin,Kalmykov:2005hb,Kalmykov:2004xg,Davydychev:2003mv,Davydychev:2000na,Davydychev:2000kw}. 

The HPLs have been implemented in the computer algebra language FORM \cite{FORM,Moch:2005uc} and in the framework of GiNaC \cite{GINAC,Vollinga:2004sn}. In a previous publication, we presented the Mathematica package {\tt HPL} \cite{Maitre:2005uu}, which implemented HPLs for real arguments. The aim of this paper is to present its extension to complex arguments. 

The paper is articulated as follows. In the first part we shortly review the properties of the harmonic polylogarithms with the emphasis on the analytical properties in the complex plane. The second part describes the second version of the Mathematica implementation {\tt HPL} of the harmonic polylogarithms that has been extended to treat arguments in the complex plane.
\section{Review of the properties of the HPLs}
In this section, we review shortly the properties of the HPLs as described in \cite{Remiddi,Maitre:2005uu}. 
\subsection{Definitions}\label{definitions}
Harmonic polylogarithms (HPL) are defined \cite{Remiddi} through recursive integration of so-called weight functions. The number of integrations is called the weight of the HPL. The usual weights are
\begin{eqnarray}
f_1(x)&=&\frac{1}{1-x},\nonumber\\
f_0(x)&=&\frac{1}{x},\nonumber\\
f_{-1}(x)&=&\frac{1}{1+x},
\end{eqnarray}
and the corresponding weight-one HPLs are
\begin{eqnarray}
H(1;x)&=&\int\limits_0^xf_1(t)\d t=\int\limits_0^x\frac{1}{1-t}\d t=-\log(1-x),\nonumber\\
H(0;x)&=&\log(x),\nonumber\\
H(-1;x)&=&\int\limits_0^xf_{-1}(t)\d t=\int\limits_0^x\frac{1}{1+t}\d t=\log(1+x).
\end{eqnarray}
HPLs of higher weights are then given by
\begin{eqnarray}
H(^n0;x)&=&\frac{1}{n!}\log^n x\nonumber\\
H(a,a_{1,\dots,k};x)&=&\int\limits_0^x f_a(t)H(a_{1,\dots,k};t)\d t\, ,
\end{eqnarray}
where we use the notations
\[^n i=\underbrace{i,...,i}_n,\quad \textnormal{and }\quad a_{1,\dots,k}=a_1,...,a_k .\]
A useful notation introduced in ref.~\cite{Remiddi} for harmonic polylogs with non-zero rightmost index is given by dropping the zeros in the vector $a$, and adding 1 to the absolute value of the next non-zero index to the right for each dropped 0. This gives for example $(3,-2)$ for $(0,0,1,0,-1)$. We can extend this notation to all index vectors by allowing zeros to take place in the rightmost position of the new index vector. This gives for example $(3,-2,0,0)$  for $(0,0,1,0,-1,0,0)$. We will enclose index vectors in this notation in curly brackets and refer to it as the ``m"-notation, as opposed to the ``a"-notation. Some formulae or transformations are more easily expressed in the one or the other notation, therefore we keep both notations in parallel.

For the extension to arguments in the complex plane it is convenient to define the linear combinations
\begin{eqnarray}
H(+;x)&=&H(1;x)+H(-1;x),\nonumber\\
H(-;x)&=&H(1;x)-H(-1;x),\nonumber\\
H(\pm,a_{1,\dots,k};x)&=&H(1,a_{1,\dots,k};t)\pm H(-1,a_{1,\dots,k};t),
\end{eqnarray}
which correspond to introducing the weight functions
\begin{eqnarray}
f_+(x)=\frac{2}{1-x^2},\qquad f_-(x)=\frac{2x}{1-x^2}\,.
\end{eqnarray}
We will refer to these two weights together with the weight $f_0$ as the ``$\pm$" weights as opposed to the ``integer" weights $f_1$, $f_0$ and $f_{-1}$.

The formula for the derivative of the HPLs follows directly from the definition,
\begin{equation}
\frac{\d}{\d x}H(a,a_{1,\dots,k};x)=f_a(x)H(a_{1,\dots,k};x).
\end{equation}
%
\subsection{Product identities and minimal set}\label{sec:algebra}
%
The product of two HPLs of same argument can be expressed as a sum of HPLs as follows,
\begin{equation}
H(p_1,\dots,p_{w_1};x)H(q_1,\dots,q_{w_2};x)=H(\mathbf{p};x)H(\mathbf{q};x)=\sum\limits_{r\in \mathbf{p}\uplus\mathbf{q}}H(\mathbf{r};x)\,,
\end{equation}
where $\mathbf{p}\uplus\mathbf{q}$ is the set of all arrangements of the elements of $\mathbf{p}$ and $\mathbf{q}$ such that the internal order of the elements of $\mathbf{p}$ and $\mathbf{q}$ is kept. For example one has for $\mathbf{p}=(a,b)$ and $\mathbf{q}=(y,z)$,
\begin{eqnarray}
H(a,b;x)H(y,z;x)&=&H(a,b,y,z;x)+H(a,y,b,z;x)\nonumber\\
&&+H(a,y,z,b;x)+H(y,a,b,z;x)\nonumber\\
&&+H(y,a,z,b;x)+H(y,z,a,b;x)\,.
\end{eqnarray}
This identity is based on the general property
\begin{equation}
\int\limits_{0}^{t}d x_1\int\limits_{0}^{t}d x_2=\int\limits_{0}^{t}d x_1\int\limits_{0}^{x_1}d x_2+\int\limits_{0}^{t}d x_2\int\limits_{0}^{x_2}d x_1\,,
\end{equation} 
and is independent of the type of weights considered.  

As a consequence, the reduction to a minimal set for HPLs of $\pm$ weights is exactly equivalent to the reduction to the minimal set for integer weights. In our implementation of the $\pm$ weights, we use exactly the same procedure as for the integer weights \cite{Maitre:2005uu}, replacing the weight $+1$ by $+$ and $-1$ by $-$.   
\subsection{Analytical properties}\label{divergences}
HPLs of integer weights have logarithmic divergences at $1,-1$ and branch cuts on $(-\infty,-1)$, $(1,\infty)$ or both depending on whether they have $1$ or $-1$ weights in their index vectors. HPLs with $\pm$ weights always have both branch cuts and have as many logarithmic singularities in both $-1$ and $1$ as there are $+$s or $-$s before the first zero from the left of the weight vector. 

HPLs have logarithmic $\log^n(0)$ divergences at 0 and a branch cut on $(-\infty,0)$ if they have $n$ 0s at the right of the weight vector.
\subsection{Argument transformation}\label{sec:transformations}
This section presents the argument transformations for HPLs. We concentrate on the transformations for HPLs of $\pm$ weights. For integer weights we refer to refs.~\cite{Remiddi,Maitre:2005uu}. Since the argument of the HPLs are now complex, we have to treat the transformations in a more detailed way. We will take the argument to lie in the complex plane with the two cuts $C_-=(-\infty,-1)$ and $C_+=(1,\infty)$. We will further denote the upper half complex plane by $\mathds{C}_+$ and the lower complex plane by $\mathds{C}_-$.  We consider the real axis on these cuts only as the limit of a complex argument approaching the real axis. So a ``real argument with infinitesimal small imaginary part" will be addressed in this section exactly the same way as a argument in the upper complex plane. 

For HPLs on the real axis the argument transformations described in ref.~\cite{Remiddi} map the integration range to a range starting or ending at a point where a potential singularity lies. Therefore these transformations  have to be applied to HPLs put into a form where the potentially divergent parts are explicitly factored out in the form of HPLs of weight one, for which the analytic continuation is known. 

Extending the HPLs to the complex plane allows to map the integration so that it avoids the divergent points for each HPLs, without having to put them in a factored form. In analogy to the ``real" transformation which introduced HPLs evaluated at unity, the ``complex" transformations introduce HPLs evaluated at $i$ and $-i$. 
\subsubsection{The transformation $x\to -x$}
We first consider the transformation $x\to -x$. It takes an argument from $\mathds{C}_+$ into $\mathds{C}_-$ and conversely. Using the definition of the HPLs,
\begin{eqnarray}
\int_0^{-x}\d x'f_a(x')H(...,x')=-\int_0^{x}\d y f_a(y)H(...,-y), 
\end{eqnarray}
we see that the weight functions transform the following way
\begin{eqnarray}
f_1(x')\to -f_{-1}(y)\,,&\quad f_0(x')\to f_0(y)\,,\quad& f_{-1}(x')\to -f_1(y)\,,\nonumber\\
f_+(x')\to -f_+(y)\,,&& f_-(x')\to +f_-(y)\;.
\end{eqnarray}
The transformation of HPLs without trailing 0s is easily done, we have for integer weights
\begin{equation}
H(n_1,...,n_w;-x)=(-1)^s H(-n_1,...,-n_w;x)\;,
\end{equation}
with $s$ given by the number of non-zero weights (the length of the vector in the ``m" notation) and for $\pm$ weights by 
\begin{equation}
H(s_1,...,s_w;-x)=(-1)^s H(s_1,...,s_w;x)\;,
\end{equation}
with this time $s$ given by the number of + weights only. So we see that HPL with $\pm$ weights and no trailing zeros are either odd or even, when the number of $+$ weights is odd or even, respectively.

The case of trailing zeros is more tricky. Consider $x=re^{i\phi}$ 
\begin{equation}
H_0(x)=\log x=\log(r)+i\phi\;,\qquad H_0(-x)=\log(-x)=\log(r)+i(\phi\pm \pi)\;,
\end{equation}
with the $+$ sign for argument in $\mathds{C_-}$ and $-$ for  $\mathds{C_+}$.
HPLs of higher weights with trailing zeros can be treated by extracting the trailing zeros using the product identities.
\subsubsection{The transformation $x\to 1/x$}
We consider the transformation 
\[x=\frac{1}{y}  \] 
for HPLs with $\pm$ weights. The results for integer weights can be obtained in the same way and are described in refs.~\cite{Remiddi,Maitre:2005uu}. The identities for weight one read
\begin{eqnarray}
H(0;x)&=&-H(0;y),\nonumber\\
H(+;x)&=&H(+;y)+ i\pi\Theta(x),\nonumber\\
H(-;x)&=&H(-;y)+2 H(0;y)+i\pi\Theta(x).
\end{eqnarray}
where $\Theta(x)$ is $+1$ in the upper half complex plane and $-1$ in the lower half complex plane.
For higher weight we proceed by induction. We take $x$ to be in the lower complex plane, so that $y$ is in the upper complex plane. We split the integration into two pieces,
\begin{eqnarray}
H(s,s_{2,\dots,k};y)&=&\int\limits_0^{1/x}\d x'f_s(x')H(s_{2,\dots,k};t)\nonumber\\
&=&\left(\int\limits_0^i\d x'+\int\limits_i^{\frac{1}{x}}\d x'\right)f_s(x')H(s_{2,\dots,k};x')\nonumber\\
&=&H(s,s_{2,\dots,k};i)-\int\limits_{-i}^{x}\frac{\d t'}{t'^2}f_s\left(\frac{1}{t'}\right)H\left(s_{2,\dots,k};\frac{1}{t'}\right)\nonumber\\
&=&H(s,s_{2,\dots,k};i)-\left(\int\limits_{0}^{x}-\int\limits_{0}^{-i}\,\right)\frac{\d t'}{t'^2}f_s\left(\frac{1}{t'}\right)H\left(s_{2,\dots,k};\frac{1}{t'}\right).
\end{eqnarray} 
We have for the different cases $s=0$ and $s=+$ and $s=-$,
\begin{eqnarray}
\frac{\d t}{t^2}f_0\left(\frac{1}{t}\right)&=&\frac{\d t}{t}\d t=f_0(t)\d t\;,\nonumber\\
\frac{\d t}{t^2}f_{+}\left(\frac{1}{t}\right)&=&\d t\left(\frac{1}{t^2-1}\right)=-f_+(t)\d t\;,\nonumber\\
\frac{\d t}{t^2}f_{-}\left(\frac{1}{t}\right)&=&\d t\left(\frac{2}{(t^2-1)t}\right)=\d t\left(\frac{2t}{t^2-1}-\frac{2}{t}\right)=\left(-f_-(t)-2 f_0(t)\right)\d t .\nonumber\\
\end{eqnarray}
We only need to replace
\[H\left(s_{2,\dots,k};\frac{1}{t'}\right)\]
by its representation in terms of HPLs of argument $t'$. The case of $x$ in the lower complex plane is treated exactly on the same way, separating the integration at $x'=-i$. 
\subsubsection{The transformation $x\to \displaystyle\frac{1-x}{1+x}$}
We consider the transformation
\[x=\frac{1-y}{1+y}\; \] 
for HPLs with $\pm$ weights. The results for the integer weights obtained on the same way and are described in \cite{Remiddi,Maitre:2005uu}. The identities for weight one read
\begin{eqnarray}
H(0;x)&=&-H(+;y),\nonumber\\
H(+;x)&=&-H(0;y),\nonumber\\
H(-;x)&=&-H(0,y)+H(+,y)-H(-;y)-2\log 2.
\end{eqnarray}
For higher weight we proceed by induction. We split the integration into two pieces,
\begin{eqnarray}
\lefteqn{H(s,s_{2\dots k};y)=\int\limits_0^{\frac{1-x}{1+x}}\d x'f_s(x')H(s_{2\dots k};t)}&&\nonumber\\
&=&\left(\int\limits_0^i\d x'+\int\limits_i^{\frac{1-x}{1+x}}\d x'\right)f_s(x')H(s_{2,\dots,k};x')\nonumber\\
&=&H(s,s_{2,\dots,k};i)-\int\limits_{-i}^{x}\frac{2\d t'  }{(1+t')^2}f_s\left(\frac{1-t'}{1+t'}\right)H\left(s_{2,\dots,k};\frac{1-t'}{1+t'}\right)\nonumber\\
&=&H(s,s_{2,\dots,k};i)-\left(\int\limits_{0}^{x}-\int\limits_{0}^{-i}\,\right)\frac{2\d t'}{(1+t')^2}f_s\left(\frac{1-t'}{1+t'}\right)H\left(s_{2,\dots,k};\frac{1-t'}{1+t'}\right)\;.\nonumber\\
\end{eqnarray} 
We have for the different cases $s=0$, $s=+$ and $s=-$,
\begin{eqnarray}
\frac{2\d t'}{(1+t')^2}f_0\left(\frac{1-t'}{1+t'}\right)&=&\frac{2\d t'}{1-t'^2}=f_+(t')\d t'\;,\nonumber\\
\frac{2\d t'}{(1+t')^2}f_{+}\left(\frac{1-t'}{1+t'}\right)&=&\frac{1}{t'}\d t'=f_0(t')\d t'\;,\nonumber\\
\frac{2\d t'}{(1+t')^2}f_{-}\left(\frac{1-t'}{1+t'}\right)&=&\d t'\frac{1-t'}{t'(1+t')}=\left(f_0(t')-f_+(t')+f_-(t')\right)\d t' .\nonumber\\
\end{eqnarray}
Now we only need to replace
\[H\left(s_{2,\dots,k};\frac{1-t'}{1+t'}\right)\]
by its expansion in terms of HPLs of argument $t'$ and perform the integration using  the definition of the HPLs. 
\subsubsection{The transformation $x^2\rightarrow x$}
This transformation is not defined for $\pm$ weights. We mention it here since it will be used for values of HPLs at $x=i$ in Section \ref{sec:MZV}. The identities for weight one are
\begin{eqnarray}
H(0;x^2)&=&\log(x^2)=2H(0;x),\nonumber\\
H(1;x^2)&=&-\log(1-x^2)=H(1;x)-H(-1;x)=H(-;x).
\end{eqnarray}
The first identity holds on the complex plane except for the branch cut $(-\infty,0)$ whereas the second equation  holds on the entire complex plane except for $(1,\infty)$. For higher weights we use the relations,
\begin{eqnarray}
H(0,m_{2\dots,k};x^2)&=&\int_0^{x^2}\frac{\d \prim{x}}{\prim{x}}H(m_{2,\dots,k};\prim{x})\nonumber\\
&=&2\int_0^{x}\frac{\d \prim{t}}{\prim{t}}H(m_{2,\dots,k};\prim{t}^2),\nonumber\\
H(1,m_{2\dots,k};x^2)&=&\int_0^{x^2}\frac{\d \prim{x}}{1-\prim{x}}H(m_{2,\dots,k};\prim{x})\nonumber\\
&=&\int_0^{x}\d \prim{t}\left(\frac{1}{1-\prim{t}}-\frac{1}{1+\prim{t}}\right)H(m_{2,\dots,k};\prim{t}^2)\nonumber\\
&=&\int_0^{x}\d \prim{t}f_-(t')H(m_{2,\dots,k};\prim{t}^2)
\end{eqnarray}
recursively, where $H(m_{2,\dots,k};\prim{t}^2)$ is expressed as HPLs of argument $\prim{t}$, which are known in a recursive approach.

\subsubsection{HPLs on the imaginary axis}
A nice property of the HPLs of weights $+$ and $-$ on the imaginary axis is that they are (provided they have no trailing zeros) either purely real or imaginary. Appending a $+$ weight to an HPL that is real on the imaginary axis will change it into purely imaginary and vice versa. Appending a $-$ weight doesn't change its real or complex nature. 
%
\subsection{Values at unity and $i$}\label{sec:MZV}
\subsubsection{Values at unity}
HPLs of argument $1$ are related to the Multiple Zeta Values (MZVs) described in the literature \cite{math.CA/9910045,Borwein:1996yq}. For positive weights, the value at unity is expressed in terms of usual MZVs and for general $m$s to colored MZVs. The relation can be found through induction and reads
\begin{eqnarray}\label{HPLMZVlink}
H(\{m_{1,\dots,k}\};1)&=&N(m_{1,\dots,k})\zeta(\tilde m_{1,\dots,k}),\quad k>1\\
H(\{m\};1)&=&\zeta(m)\,,\qquad\qquad \qquad \qquad m>0\\
H(\{-m\};1)&=&(1-2^{1-m})\zeta(m)\qquad\qquad m>0\, ,
\end{eqnarray}
where the MZVs $\zeta$ are defined by
\begin{equation}\label{MZVdef}
\zeta(m_1,\dots,m_k)=\sum\limits_{i_1=1}^\infty\sum\limits_{i_2=1}^{i_1-1}\dots\sum\limits_{i_k=1}^{i_{k-1}-1}\prod\limits_{j=1}^k\frac{\sgn (m_j)^{i_j}}{i_j^{|m_j|}}\, .
\end{equation}
The vector $\tilde m$ is obtained from the vector $m$ via
\begin{equation}
\tilde m=(m_1,\sgn(m_1)m_2,\dots, \sgn(m_{i-1}) m_i,\dots ,\sgn(m_{k-1}) m_k).
\end{equation}
The factor $N(m_{1,\dots,k})$ is given by
\begin{equation}
N(m_{1,\dots,k})=(-1)^{\#(m_i<0)}.
\end{equation}
 The MZVs also form an algebra. Due to this fact, they can be expressed in terms of a few mathematical constants like powers of $\pi$, $\zeta$-functions and polylogs at specified values. We list some of the identities of ref.~\cite{math.CA/9910045,Borwein:1996yq} in Appendix \ref{sec:MZVtable}.
For the implementation of the HPL at unity, we translated the tables of the FORM package {\tt harmpol.h} \cite{Remiddi} and their expansions for weight 7 and 8 {\tt htable7.prc} and {\tt htable8.prc} to Mathematica.  

In these tables, there appear some constants that are not expressible through known constants like $\pi$, $\zeta(n)$, $\log(2)$, or $Li_n(1/2)$. Using the different relations between the different MZVs, one can reduce the number of independent constants. The list of these independent constants are listed in Appendix \ref{MZVindep}.

For index vectors with leftmost 1s, there appear divergences when the argument approaches unity. Using the relations between HPLs and harmonic sums\footnote{see section 5 of ref.\cite{Remiddi}.}, these divergences can be regulated in a proper way and expressed in terms of one divergent symbol
\[\sum\limits_{i=1}^{\infty}\frac{1}{i}=S(1,\infty)=H(1;1).\]

Care has to be taken when taking the limit $x\to 1$ in expressions containing such divergences. Since we have, for example 
\begin{eqnarray}
H(1,1;x)=\frac{1}{2}H(1;1)^2\,,
\end{eqnarray}
but
\begin{eqnarray}
H(1,1;1)=\frac{1}{2}H(1;1)^2-\frac{1}{12}\pi^2\,.
\end{eqnarray}
we see that taking the limit $x\to 1$ does not commute with the product identity. One should therefore express all the HPLs divergent in the limit $x\to 1$ in the same form (i.e. either with the logarithmic divergences factored out using the product identities, or with products of HPLs expanded in terms of a sum of HPLs of higher weight) and only then take the limit $x\to 1$.  
\subsubsection{Values at $i$}
The argument transformations of Section \ref{sec:transformations} use the values of the HPLs with $\pm$ weights at $x=i$. 
For the lower weights we have
\begin{eqnarray}
H(+,i)=\frac{i\pi}{2}\;,\rule{2.6cm}{0cm}
&&
H(-,i)=\log(2)\;,\\
H(+,-,i)=i \left(2 \mathcal{C}-\pi  \log (2)\right),
&&
H(-,+,i)=i \left(-2 \mathcal{C}+\frac{1}{2} \pi  \log (2)\right),\qquad\\
H(0,+,i)=2 i  {\mathcal C}\;,\rule{2.2cm}{0cm}
&&
H(0,-,i)=-\frac{\pi}{24}\;,\\
H(+,0,i)=  -2 i \mathcal{C} -\frac{\pi^2}{4}\;,\rule{0.9cm}{0cm}
&& H(-,0,i)= -\frac{\pi}{24} -\frac{i}{2} \pi  \log (2)\;.
\end{eqnarray}

where $ \mathcal C$ is Catalan's constant,
\[{\mathcal C}=\sum\limits_{k=0}^\infty \frac{1}{(2 k+1)^2}.\]
To find the values at $i$ for higher weights, one can use the following properties.
\begin{itemize}
\item The product identities relate HPLs in $i$ with HPLs of lower weights at $i$.
\item Some HPLs can be written in terms of more commonly known functions, whose values at $x=i$ are known.
\item Using the argument transformation $x^2\to x$ one can relate the values of HPLs with only ``-" and 0 weights to the values of HPLs with integer weights at $-1$.
\item Since the ``complex" transformations of section \ref{sec:transformations} involve HPLs at $i$ and the ``real" ones not, comparing them leads to relations between different HPLs at $i$. 
\end{itemize}  
The number of relations is, however not sufficient to solve for all HPLs at $i$. A list of values of HPLs at $i$ is given in Appendix \ref{sec:HPLItable}.

\section{The Mathematica implementation \HPL\tt\, 2}
In this section, we describe the extension of the Mathematica implementation {\tt HPL} to complex arguments. Version 2 of {\tt HPL} includes all features of the first version as well as the following new elements,
\begin{itemize}
\item numerical evaluation of HPLs for complex arguments,
\item new weights $+$ and $-$,
\item new function {\tt HPLArgTransform},
\item new integration procedure {\tt HPLInt}.
\end{itemize}
For completeness, we will not restrict the description to the new features of {\tt HPL} but will also repeat the features that did not change.  
 The package can be found at link \cite{HPLhomepage} where installation instructions can be found. After installation, the package can be loaded via
\vspace{0.2cm}\\
\rule{-1cm}{0cm}\fbox{\parbox{\exampleboxlength}{
\example\\
\example\\
\example\\
\example\\
\example\\
\example\\
\example\\
\example
}
}\vspace{0.1cm}\\
\noindent It should be loaded at the beginning of the {\tt Mathematica} session. 
\subsection{New functions}\label{sec:newfunctions}
The package {\tt HPL} defines the following new functions.
\begin{itemize}
\item {\tt HPL[m,x]} is the harmonic polylogarithm $H(m;x)$, where {\tt m} is a list representing the index vector. For integer weights, we chose the ``m"-notation as the standard notation. It is possible to give as the argument a vector in the ``a"-notation, or even in a mix between the two notations, as no confusion is possible. Results will be displayed in the ``m"-notation. It is also possible to mix $\pm$ and integer weights.   
\vspace{0.2cm}\\
\rule{-1cm}{0cm}\fbox{\parbox{\exampleboxlength}{
\example\\
\example\\
\example\\
\example\\
\example
}
}\vspace{0.1cm}\\
\item {\tt HPLMtoA[m\_List]} and {\tt HPLAtoM[a\_List]} convert vectors from the ``m"- to the ``a"-, and from the ``a"- to the ``m"-notation respectively. Both can convert vectors which mix the two notations.
\vspace{0.2cm}\\
\rule{-1cm}{0cm}\fbox{
\parbox{\exampleboxlength}{
\example\\
\example\\
\example\\
\example\\
\example\\
\example
}
}\vspace{0.1cm}\\
\item {\tt HPLpm21m1} and {\tt HPL1m12pm} convert HPLs with $\pm$ weights into HPLs with integer weights and vice versa.
\vspace{0.2cm}\\
\rule{-1cm}{0cm}\fbox{\parbox{\exampleboxlength}{
\example\\
\example\\
\example\\
\example
}
}\vspace{0.1cm}\\
It is also possible to convert a function with integer weights into the corresponding sum of functions of $\pm$ weights, and vice versa.
\vspace{0.2cm}\\
\rule{-1cm}{0cm}\fbox{\parbox{\exampleboxlength}{
\example\\
\example\\
\example\\
\example
}
}\vspace{0.1cm}\\
\item {\tt HPLLogExtract} extracts the singular behavior of HPLs using the product identities. The logarithmic divergences are extracted  for
\begin{description}
\item[integer weights] in their argument at 0 and 1, 
\item[$+-$ weights] in their argument only at 0. 
\end{description}
The result is displayed as function of $\log(x),\log(1-x)$ or $H(1;x),H(0;x)$ depending on the option settings (see Section \ref{sec:options}).
\vspace{0.2cm}\\
\rule{-1cm}{0cm}\fbox{\parbox{\exampleboxlength}{
\example\\
\example\\
\example\\
\example\\
\example\\
\example
}
}\vspace{0.1cm}\\
In the case of mixed integer and $\pm$ weights, only the divergences at 0 are factored.
\item {\tt HPLConvertToKnownFunctions} returns its argument with HPLs replaced by their representation in terms of more common functions, whenever possible. 
\vspace{0.2cm}\\
\rule{-1cm}{0cm}\fbox{\parbox{\exampleboxlength}{
\example\\
\example\\
\example\\
\example\\
\example\\
\example
}
}\vspace{0.1cm}\\
It is only needed if the option {\tt \$HPLAutoConvertToKnownFunctions} is set to {\tt False} (see Section \ref{sec:options}).
\item {\tt HPLProductExpand} returns the value obtained by replacing products of HPLs of same argument by their representation as a linear combination of HPLs, as presented in Section \ref{sec:algebra}.  
\vspace{0.2cm}\\
\rule{-1cm}{0cm}\fbox{\parbox{\exampleboxlength}{
\example\\
\example\\
\example\\
\example\\
\example\\
\example
}
}\vspace{0.1cm}\\
In order to expand all products, {\tt HPLProductExpand} expands its argument (using {\tt Expand}), so that terms of the form 
\[H(\dots;x)\big(H(\dots;x)+H(\dots;x)+\dots\big).\]
are also replaced. For large expressions this might be time consuming, one should in this case first collect the products of HPLs and apply {\tt HPLProductExpand} only to the products and not to the whole expression. 
\vspace{0.2cm}\\
\rule{-1cm}{0cm}\fbox{\parbox{\exampleboxlength}{
\example\\
\example
}
}\vspace{0.1cm}\\
 Since the product properties are general, the function works for any vectors of symbolic weights.
\vspace{0.2cm}\\
\rule{-1cm}{0cm}\fbox{\parbox{\exampleboxlength}{
\example\\
\example
}
}\vspace{0.1cm}\\
 
\item {\tt HPLConvertToSimplerArgument} returns its argument with HPLs of related arguments replaced by their expansion as a sum of HPLs of simpler arguments. Which transformations are implemented depends on the type of  weights. They are: 
\begin{description}
\item[integer weights] $-x$, $x^2$, $1-x$, $1/x$, $x/(x-1)$ and $(1-x)/(1+x)$ 
\item[$\pm$ weights] $-x$, $1/x$, and $(1-x)/(1+x)$. 
\end{description}
\rule{10cm}{0cm}\vspace{0.2cm}\\
\rule{-1cm}{0cm}\fbox{\parbox{\exampleboxlength}{
\example\\
\example\\
\example\\
\example\\
\example\\
\example\\
\example\\
\example\\
\example\\
\example
}
}\vspace{0.1cm}\\
The transformations do not work for mixed integer and $\pm$ weights.
\item {\tt HPLArgTransform[x ,r ,$\delta$ ]} returns its first argument expressed in terms of HPLs of the  transformed argument. It does the same job as {\tt HPLConvertToSimplerArgument}, but the transformation to be performed is not fixed by the form of the argument of the HPL but is specified as an argument of {\tt HPLArgTransform}. 
\vspace{0.2cm}\\
\rule{-1cm}{0cm}\fbox{\parbox{\exampleboxlength}{
\example\\
\example\\
\example\\
\example\\
\example\\
\example\\
\example
}
}\vspace{0.1cm}\\
In addition one can specify whether the argument is assumed to lie in the upper ($\delta$=+1) or lower ($\delta=-1$) complex plane, or whether to leave it purely  symbolic.
\vspace{0.1cm}
\vspace{0.2cm}\\
\rule{-1cm}{0cm}\fbox{\parbox{\exampleboxlength}{
\example\\
\example\\
\example\\
\example\\
\example\\
\example\\
\example\\
\example\\
\example\\
\example
}
}\vspace{0.1cm}
 The transformations available are the same as for {\tt HPLConvertToSimplerArgument}, with the same restrictions on the weight vector. 
\vspace{0.2cm}\\
\rule{-1cm}{0cm}\fbox{\parbox{\exampleboxlength}{
\example\\
\example\\
\example\\
\example\\
\example\\
\example\\
\example\\
\example\\
\example
}
}\vspace{0.1cm}
 
The rule {\tt r} can be written using any symbol, which makes the code more intuitive. The symbol in the rule does not have to match the arguments of the HPLs.   
\vspace{0.2cm}\\
\rule{-1cm}{0cm}\fbox{\parbox{\exampleboxlength}{
\example\\
\example\\
\example\\
\example
}
}\vspace{0.1cm}
\item {\tt HPLReduceToMinimalSet} returns its argument with the HPLs projected to the minimal set, as described in Section \ref{sec:algebra}. 
\vspace{0.2cm}\\
\rule{-1cm}{0cm}\fbox{\parbox{\exampleboxlength}{
\example\\
\example\\
\example\\
\example
}
}\vspace{0.1cm}
\item {\tt HPLAnalyticContinuation[x,HPLAnalyticContinuationRegion->region]} assumes that the arguments of the HPLs are  real and returns its argument {\tt x} with HPLs replaced by their analytic continuation.
The arguments of the HPLs are taken to belong to the interval specified by the option  {\tt HPLAnalyticContinuationRegion}, for which {\tt region} can be either
\begin{description}
\item[{\tt minftom1}] the interval $-\infty$ to $-1$
\item[{\tt m1to0}] the interval $-1$ to $0$
\item[{\tt onetoinf}] the interval $1$ to $\infty$.
\end{description}
The HPLs are replaced by their representation in terms of HPLs of argument in the interval $(0,1)$. The choice of the side of the branch cut from which the argument is approached is set by the option 

{\tt HPLAnalyticContinuationSign}\\ 
which can take values $-1$, $1$ or any symbol.
\vspace{0.2cm}\\
\rule{-1cm}{0cm}\fbox{\parbox{\exampleboxlength}{
\example\\
\example\\
\example\\
\example
}
}\vspace{0.1cm}\\
If the option {\tt HPLAnalyticContinuationRegion} is omitted, and if the argument is numerical, {\tt HPLAnalyticContinuation} will automatically use the appropriate setting. If the option {\tt HPLAnalyticContinuationSign} is omitted, {\tt HPLAnalyticContinuation} will use the value stored in the variable {\tt \$HPLAnalyticContinuationSign} which is set by default to $1$\footnote{This is the same convention as ref.~\cite{Gehrmann:2001pz}, but opposite to that of ref.~\cite{Vollinga:2004sn}}. 
\vspace{0.2cm}\\
\rule{-1cm}{0cm}\fbox{\parbox{\exampleboxlength}{
\example\\
\example\\
\example\\
\example
}
}\vspace{0.1cm}\\
It is to be noted that the {\tt Mathematica} conventions for the analytic continuation are not always the same as that of the {\tt HPL} package. This is illustrated by the following example
\vspace{0.2cm}\\
\rule{-1cm}{0cm}\fbox{\parbox{\exampleboxlength}{
\example\\
\example\\
\example\\
\example
}
}\vspace{0.1cm}\\
Here {\tt HPL} takes the argument of the HPL to have an infinitesimal positive imaginary part, whereas Mathematica take the argument of the logarithm to have a positive imaginary part. Since the substitution of HPLs through more common functions has precedence over the analytic continuation, the option 

{\tt \$HPLAutoConvertToKnownFunctions} 
\\
can interfere with the analytic continuation.
\vspace{0.2cm}\\
\rule{-1cm}{0cm}\fbox{\parbox{\exampleboxlength}{
\example\\
\example\\
\example\\
\example\\
\example\\
\example
}
}\vspace{0.1cm}\\
This example shows that with the option 

{\tt \$HPLAutoConvertToKnownFunctions} 

set to {\tt True} we lose control over the sign of the imaginary part (as in the first case it will now depend on {\tt Mathematica}'s conventions).
\item {\tt HPLInt[x,t]} returns a primitive of {\tt x} with respect to {\tt t}.  The cases working are for $k>0$
\begin{eqnarray*}
\int d t \frac{C t^j}{(1-t^2)^k}H(...,t), && 
\int d t \frac{C t^j}{(1-t)^k}H(...,t), \\ 
\int d t \frac{C t^j}{(1+t)^k}H(...,t), && 
\int d t \,C t^j H(...,t).  
\end{eqnarray*}
\vspace{0.2cm}\\
\rule{-1cm}{0cm}\fbox{\parbox{\exampleboxlength}{
\example\\
\example\\
\example\\
\example\\
\example\\
\example
}
}\vspace{0.1cm}

\item {\tt MZV[m]} is the Multiple Zeta Value (see Section \ref{sec:MZV}) corresponding to the index vector $m$. Their value in terms of mathematical constants are tabulated\footnote{these tables are those of the FORM package {\tt harmpol}\cite{Vermaseren_harmpol}.} up to weight 8 and systematically replaced. For higher weights, the cases covered by Appendix \ref{sec:MZVtable} are also replaced.
\vspace{0.2cm}\\
\rule{-1cm}{0cm}\fbox{\parbox{\exampleboxlength}{
\example\\
\example\\
\example\\
\example\\
\example\\
\example
}
}\vspace{0.1cm}
\item {\tt HPLI[m]} is the HPL of weight vector {\tt m} evaluated at argument $i$. Their values in terms of mathematical constants are tabulated up to weight 8 and systematically replaced. For higher weights, the cases covered by Appendix \ref{sec:HPLItable} are also replaced. 
 \vspace{0.2cm}\\
 \rule{-1cm}{0cm}\fbox{\parbox{\exampleboxlength}{
 \example\\
 \example\\
 \example\\
 \example
 }}\vspace{0.1cm}
\item The function {\tt \$HPLOptions} gives a list of the options of the package and their current values.
\vspace{0.2cm}\\
\rule{-1cm}{0cm}\fbox{\parbox{\exampleboxlength}{
\example\\
\example
}
}
\item The variable {\tt \$HPLFunctions} contains a list of the functions provided by the package.
\vspace{0.2cm}\\
\rule{-1cm}{0cm}\fbox{\parbox{\exampleboxlength}{
\example\\
\example
}
}
\end{itemize}
\subsection{Functions modified}
\begin{itemize}
\item We define the derivatives of HPLs as described in Section \ref{definitions}. 
\vspace{0.2cm}\\
\rule{-1cm}{0cm}\fbox{\parbox{\exampleboxlength}{
\example\\
\example\\
\example\\
\example
}
}\vspace{0.1cm}
The integration showing up in the recursive definition of the HPLs is also implemented. It is however recommended to use the function {\tt HPLInt}.
\vspace{0.2cm}\\
\rule{-1cm}{0cm}\fbox{\parbox{\exampleboxlength}{
\example\\
\example\\
\example\\
\example
}
}\vspace{0.1cm}
\item The function {\tt Series} is able to expand HPLs around $x=0$ and $x=1$.
\vspace{0.2cm}\\
\rule{-1cm}{0cm}\fbox{\parbox{1.1\textwidth}{
\example\\
\example\\
\example\\
\example\\
\example\\
\example\\
\example\\
\example
}
}\end{itemize}
\subsection{Working with the options}\label{sec:options}
The package {\tt HPL} has some options to control its behavior. They set the preferred form in which expressions are displayed. The options can be overridden locally by the functions described above. The effects of the options are described in the following.
\begin{description}
\item[$\bullet$ {\tt \$HPLAutoConvertToKnownFunctions}:] If set to {\tt True}, HPLs will be converted to more common functions (logs, polylogarithms, Nielsen polylogs) if possible, using the identities of Appendix \ref{sec:identities}. This might be counterproductive when the properties of the HPLs are more explicit in the HPL form than other representations.
\vspace{0.2cm}\\
\rule{-1cm}{0cm}\fbox{\parbox{\exampleboxlength}{
\example\\
\example\\
\example\\
\example\\
\example\\
\example\\
\example\\
\example
}
}\vspace{0.1cm}\\
Furthermore, if this options is set to {\tt True} while using the analytic continuation described above, the result may be wrong, as {\tt Mathematica} does not have different conventions for the analytic continuation. Default is {\tt False}.
\item[$\bullet$ {\tt \$HPLAutoProductExpand}:] If {\tt True} the products of HPLs are automatically converted into a sum of HPL of weight equal to the sum of the weights of the two factors, as described in section \ref{sec:algebra}. Default is {\tt False}.  
\vspace{0.2cm}\\
\rule{-1cm}{0cm}\fbox{\parbox{\exampleboxlength}{
\example\\
\example\\
\example\\
\example\\
\example\\
\example
}
}\vspace{0.1cm}\\
Setting the option {\tt \$HPLAutoProductExpand} {\tt True} affects only explicit products and does not act on factorized products like the function {\tt HPLProductExpand}. 
\vspace{0.2cm}\\
\rule{-1cm}{0cm}\fbox{\parbox{\exampleboxlength}{
\example\\
\example\\
\example\\
\example
}
}\vspace{0.1cm}
\item[$\bullet$  {\tt \$HPLAutoLogExtract}:] If {\tt True} the logarithmic divergences  $\log(1-x)$ and $\log(x)$ are automatically extracted from the HPLs.
The default setting is {\tt False}.  
\vspace{0.2cm}\\
\rule{-1cm}{0cm}\fbox{\parbox{\exampleboxlength}{
\example\\
\example\\
\example\\
\example\\
\example\\
\example\\
\example\\
\example
}
}\vspace{0.1cm}\\
The extraction of the divergent behavior only makes sense, if one does not re-expand the products automatically with the option {\tt \$HPLAutoProductExpand} set to {\tt True}. If the latter option is set to {\tt True}, the option {\tt \$HPLAutoLogExtract} will have no effect.
\vspace{0.2cm}\\
\rule{-1cm}{0cm}\fbox{\parbox{\exampleboxlength}{
\example\\
\example\\
\example\\
\example\\
\example\\
\example\\
\example
}
}\vspace{0.1cm}\\
On the other hand, if the option {\tt \$HPLAutoConvertToKnownFunctions} is set to {\tt True}, the out factorized HPLs of weight one will be replaced by logs before being re-expanded, as shown by the following example.
\vspace{0.2cm}\\
\rule{-1cm}{0cm}\fbox{\parbox{\exampleboxlength}{
\example\\
\example\\
\example\\
\example\\
\example
}
}\vspace{0.1cm}
\item[$\bullet$ {\tt \$HPLAutoReduceToMinimalSet}:] If set to {\tt True}, the HPLs will be automatically reduced to a minimal basis (up to weight 8). This only makes sense if one does not expand the obtained products again, or if the factors of smaller weight can be replaced by their expression in terms of known functions. Therefore, for the reduction to be performed, one has to have the option 

{\tt \$HPLAutoProductExpand} equal to {\tt False} or 

{\tt \$HPLAutoConvertToKnownFunctions} equal to {\tt True}. 
\\
If this is not fulfilled, the option will have no effect. It defaults to {\tt False}.
\vspace{0.2cm}\\
\rule{-1cm}{0cm}\fbox{\parbox{\exampleboxlength}{
\example\\
\example\\
\example\\
\example\\
\example\\
\example\\
\example\\
\example
}
}\vspace{0.1cm}
\item[$\bullet$ {\tt \$HPLAutoConvertToSimplerArgument}:] If set to {\tt True}, HPLs of arguments $-x$, $x^2$, $1-x$, $1/x$, $x/(x-1)$ and $(1-x)/(1+x)$ will be automatically replaced by their representation in term of HPLs of argument $x$ along the lines of Section \ref{sec:transformations}. Its default is {\tt False}.
\vspace{0.2cm}\\
\rule{-1cm}{0cm}\fbox{\parbox{\exampleboxlength}{
\example\\
\example\\
\example\\
\example\\
\example\\
\example\\
\example
}
}\vspace{0.1cm}
\item[$\bullet$ {\tt HPLAnalyticContinuationSign}:] If set to 1 the analytic continuation of the HPLs is taken assuming a positive infinitesimal imaginary part for arguments, if set to $-1$ a negative one is assumed. {\tt \$HPLAnalyticContinuationSign} is only the default setting and can be overridden by specifying the option {\tt AnalyticContinuationSign} in the function {\tt HPLAnalyticContinuation}, as described above. {\tt \$HPLAnalyticContinuationSign} can be a symbol. The default setting is $+1$.
\vspace{0.2cm}\\
\rule{-1cm}{0cm}\fbox{\parbox{\exampleboxlength}{
\example\\
\example\\
\example\\
\example\\
\example\\
\example
}
}\vspace{0.1cm}

\end{description}

\subsection{Numerical evaluation}
The package {\tt HPL} 2 provides an arbitrary-precision numerical evaluation in the whole complex plane.
\subsubsection{Real argument} 
HPLs of real arguments with finite precision are automatically evaluated to the precision of the argument.
\vspace{0.2cm}\\
\rule{-1cm}{0cm}\fbox{\parbox{\exampleboxlength}{
\example\\
\example\\
\example\\
\example\\
\example\\
\example
}
}\vspace{0.1cm}\\ 
For arguments with infinite precision, numerical evaluation is only undertaken if explicitly requested. 
\vspace{0.2cm}\\
\rule{-1cm}{0cm}\fbox{\parbox{\exampleboxlength}{
\example\\
\example\\
\example\\
\example\\
\example\\
\example
}
}\vspace{0.1cm}\\ 

For the numerical values of the MZV, we implemented the procedure described in refs.~\cite{math.CA/9910045,Vollinga:2004sn}.  
\vspace{0.2cm}\\
\rule{-1cm}{0cm}\fbox{\parbox{\exampleboxlength}{
\example\\
\example\\
\example\\
\example\\
\example\\
\example\\
\example\\
\example\\
\example\\
\example
}
}\vspace{0.2cm}\\
For real arguments outside the interval $[0,1]$, one can specify the sign of the infinitesimal imaginary part to be used for the analytic continuation  by appending the option {\tt AnalyticContinuationSign} as for the function {\tt HPLAnalyticContinuation}.
\vspace{0.2cm}\\
\rule{-1cm}{0cm}\fbox{\parbox{\exampleboxlength}{
\example\\
\example\\
\example\\
\example\\
\example\\
\example
}
}

\subsubsection{Complex argument}

\begin{figure}[h]
\begin{center}
\includegraphics[scale=0.8]{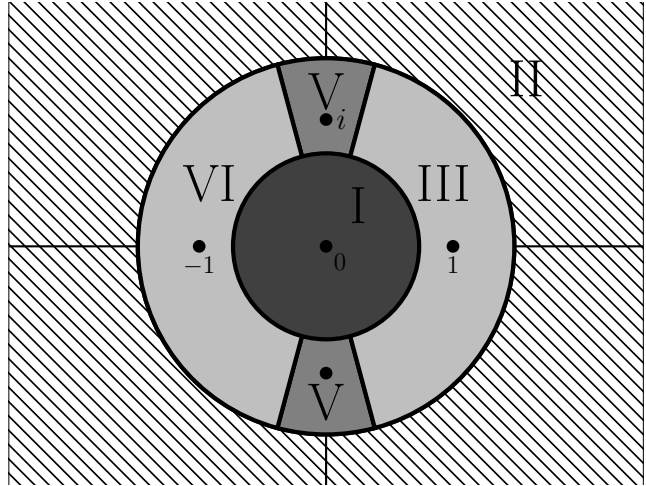}
\end{center}
\caption{Complex plane division for the numerical evaluation}
\label{fig:regions}
\end{figure}

For the sake of the numerical evaluation for complex arguments, we divide the complex plane into five different regions as pictured by figure \ref{fig:regions}.
\begin{description}
\item[region I: $|z|<0.9$] \rule{1cm}{0cm}\\
the series expansion is used
\item[region II: $|z|>1.5$] \rule{1cm}{0cm}\\
First the argument is brought into region I using the argument transformation $z\to 1/z$ described in Section \ref{sec:transformations}, then the HPLs are evaluated using series expansions.
\item[region III: $0.9<|z|<1.5$ and $|\arg(z)|<5\pi/12$] \rule{1cm}{0cm}\\
First the argument is brought into region I using the argument transformation $z\to (1-z)/(1+z)$ described in Section \ref{sec:transformations}, then the HPLs are evaluated using series expansions.
\item[region IV: $0.9<|z|<1.5$ and $|\arg(z)|>7\pi/12$] \rule{1cm}{0cm}\\
First the argument is brought into region I using the argument transformations $z\to -z$ and $z\to (1-z)/(1+z)$ described in Section \ref{sec:transformations}, then the HPLs are evaluated using series expansions.
\item[region V: $0.9<|z|<1.5$ and $5\pi/12<|\arg(z)|<7\pi/12$] \rule{1cm}{0cm}\\
The evaluation is done using the H\"older convolution described in ref.~\cite{math.CA/9910045}.
\end{description} 

The transformations mentioned above are performed using the ``complex" argument transformation for the $\pm$ weights and the ``real" transformations for integer weights. The evaluation for the $\pm$ weights in region V is done by first converting to the integer weights and then using the H\"older convolution.  

HPLs with numerical complex arguments with finite precision are evaluated automatically.
\vspace{0.2cm}\\
\rule{-1cm}{0cm}\fbox{\parbox{\exampleboxlength}{
\example\\
\example\\
\example
}
}\vspace{0.2cm}\\
HPLs with complex argument with infinite precision are only evaluated if explicitly requested.
\vspace{0.2cm}\\
\rule{-1cm}{0cm}\fbox{\parbox{\exampleboxlength}{
\example\\
\example\\
\example\\
\example\\
\example\\
\example\\
\example
}
}\vspace{0.2cm}\\
Thanks the possibility of evaluating the HPLs numerically, {\tt Plot} is able to represent HPLs graphically on the real axis
\vspace{0.2cm}\\
\rule{-1cm}{0cm}\fbox{\parbox{\exampleboxlength}{
\example\\
\example
}
}\vspace{0.1cm}\\
and in the complex plane, here are two examples of HPLs plotted on the interval $[-5,5]$ and in the unit circle\vspace{0.2cm}\\
\rule{-1cm}{0cm}\fbox{\parbox{\exampleboxlength}{
\example\\
\example\\
\example
}
}\vspace{0.2cm}\\
\rule{-1cm}{0cm}\fbox{\parbox{\exampleboxlength}{
\example\\
\example\\
\example
}
}\vspace{0.1cm}\\
We checked numerical agreement with ref.~\cite{Gehrmann:2001pz} at double precision accuracy. The values of the MZV have been checked against those of the {\tt EZface} application\cite{EZface} and the GiNaC implementation \cite{Vollinga:2004sn}. Concerning speed, our implementation cannot be expected to compete with the GiNaC implementation.

\section{Conclusion}

In this paper we presented the extension of the existing package HPL to deal with complex argument. The new version of the package provides the numerical evaluation of HPLs of complex argument over the complete complex plane. We also presented new weights more appropriate for complex arguments. They will be used for the expansion of hypergeometrical functions around half integer parameters \cite{hypexp2}. 

\section*{Acknowledgement}
We would like to thank Thomas Gehrmann for many useful discussions, David Kosower and Thomas Gehrmann for valuable comments on the manuscript and the Swiss National Science Foundation (SNF) which supported this work under contract 200020-109162.
\appendix
\section{HPLs at $i$}\label{sec:HPLItable}
In these tables, we collect the values for the HPLs at $i$ up to weight 3. 
\subsection{Weight 1}
\begin{eqnarray}
H(1;i)=\frac{i \pi }{4}-\frac{\log (2)}{2}\;,& &   H(-1;i)= \frac{i \pi }{4}+\frac{\log (4)}{4}\;,\\
H(+;i)=\frac{i\pi}{2}  \;,                    & &   H(-;i)= -\log (2)\;,\\
H(0;i)=\frac{i\pi}{2}\;.&&
\end{eqnarray}

\subsection{Weight 2}
\begin{eqnarray}
H(+,i)=\frac{i\pi}{2}\;,\rule{2.6cm}{0cm}
&&
H(-,i)=\log(2)\;,\\
H(+,-,i)=i \left(2 \mathcal{C}-\pi  \log (2)\right)\;,
&&
H(-,+,i)=i \left(-2 \mathcal{C}+\frac{1}{2} \pi  \log (2)\right),\qquad\\
H(0,+,i)=2 i  {\mathcal C}\;,\rule{2.2cm}{0cm}
&&
H(0,-,i)=-\frac{\pi}{24}\;,\\
H(+,0,i)=  -2 i \mathcal{C} -\frac{\pi^2}{4}\;,\rule{0.9cm}{0cm}
&& H(-,0,i)=- \frac{\pi}{24} -\frac{i}{2} \pi  \log (2)\;.
\end{eqnarray}
\subsection{Weight 3}
\begin{eqnarray}
H(+,+,+;i)&=&-\frac{i \pi ^3}{48}\;,\\
 H(+,+,-;i)&=&\frac{1}{16} \left(\pi ^2 \log (16)-21 \zeta (3)\right)\;,\\
 H(+,+,0;i)&=&-\frac{i \pi ^3}{16}+\frac{7 \zeta (3)}{4}\;,\\
 H(+,-,+;i)&=&-{\mathcal C} \pi +\frac{21 \zeta (3)}{8}\;,\\
H(+,-,-;i)&=&-\frac{7 i \pi ^3}{96}-2 i {\mathcal C} \log (2)+\frac{7}{8} i \pi  \log ^2(2)-2 \mathrm{Li}_3\left(\frac{1}{2}-\frac{i}{2}\right)\nonumber\\
 &&+2 \text{Li}_3\left(\frac{1}{2}+\frac{i}{2}\right)\;,\\
H(+,-,0;i)&=&-{\mathcal C} \pi +\frac{7 i \pi ^3}{48}+\frac{1}{2} \pi ^2 \log (2)+\frac{1}{4} i \pi  \log ^2(2)+4 \text{Li}_3\left(\frac{1}{2}-\frac{i}{2}\right)\nonumber\\
&&-4 \text{Li}_3\left(\frac{1}{2}+\frac{i}{2}\right)\;.
 \end{eqnarray}
\begin{eqnarray}
 H(+,0,+;i)&=&{\mathcal C} \pi -\frac{7 \zeta (3)}{2}\;,\\
 H(+,0,-;i)&=&-\frac{11 i \pi ^3}{96}+2 i {\mathcal C} \log (2)-\frac{1}{8} i \pi  \log ^2(2)-2 \text{Li}_3\left(\frac{1}{2}-\frac{i}{2}\right)\nonumber\\
 &&+2 \text{Li}_3\left(\frac{1}{2}+\frac{i}{2}\right)\;,\\
 H(+,0,0;i)&=&{\mathcal C} \pi\;, \\
 H(-,+,+;i)&=&{\mathcal C} \pi -\frac{1}{8} \pi ^2 \log (2)-\frac{21 \zeta (3)}{16}\;,\\
 H(-,+,-;i)&=&\frac{7 i \pi ^3}{48}+2 i {\mathcal C} \log (2)-\frac{3}{4} i \pi  \log ^2(2)+4 \text{Li}_3\left(\frac{1}{2}-\frac{i}{2}\right)\nonumber\\
 &&-4 \text{Li}_3\left(\frac{1}{2}+\frac{i}{2}\right)\;,\\
 H(-,+,0;i)&=&{\mathcal C} \pi -\frac{i \pi ^3}{32}-\frac{1}{4} \pi ^2 \log (2)-\frac{1}{8} i \pi  \log ^2(2)-2 \text{Li}_3\left(\frac{1}{2}-\frac{i}{2}\right)\nonumber\\
 &&+2 \text{Li}_3\left(\frac{1}{2}+\frac{i}{2}\right)\;,\\
 H(-,-,+;i)&=&-\frac{7 i \pi ^3}{96}+\frac{1}{8} i \pi  \log ^2(2)-2 \text{Li}_3\left(\frac{1}{2}-\frac{i}{2}\right)\nonumber\\
 &&+2 \text{Li}_3\left(\frac{1}{2}+\frac{i}{2}\right)\;,\\
 H(-,-,-;i)&=&-\frac{1}{6} \log ^3(2)\;,\\
 H(-,-,0;i)&=&\frac{1}{48} \left(12 i \pi  \log ^2(2)-\pi ^2 \log (4)+3 \zeta (3)\right)\;,\\
 H(-,0,+;i)&=&-\frac{3 i \pi ^3}{32}-\frac{1}{8} i \pi  \log ^2(2)-2 \text{Li}_3\left(\frac{1}{2}-\frac{i}{2}\right)\nonumber\\
 &&+2 \text{Li}_3\left(\frac{1}{2}+\frac{i}{2}\right)\;,\\
 H(-,0,-;i)&=&\frac{1}{24} \left(\pi ^2 \log (2)-3 \zeta (3)\right)\;,\\
 H(-,0,0;i)&=&\frac{1}{48} i \left(\pi ^3-6 i \pi ^2 \log (2)+9 i \zeta (3)\right)\;,\\
 H(0,+,+;i)&=&-{\mathcal C} \pi +\frac{7 \zeta (3)}{4}\;,\\
 H(0,+,-;i)&=&-\frac{1}{32} i \left(\pi ^3+64 {\mathcal C} \log (2)+4 \pi  \log ^2(2)\right)-2 \text{Li}_3\left(\frac{1}{2}-\frac{i}{2}\right)\nonumber\\
 &&+2 \text{Li}_3\left(\frac{1}{2}+\frac{i}{2}\right)\;,\\
 H(0,+,0;i)&=&-{\mathcal C} \pi -\frac{i \pi ^3}{8}\;.
 \end{eqnarray}
 \begin{eqnarray}
 H(0,-,+;i)&=&\frac{i}{8} \left(\pi ^3+2 \pi  \log ^2(2)\right)+4 \text{Li}_3\left(\frac{1}{2}-\frac{i}{2}\right)\nonumber\\
 &&-4 \text{Li}_3\left(\frac{1}{2}+\frac{i}{2}\right)\;,\\
 H(0,-,-;i)&=&\frac{\zeta (3)}{16}\;,\\
 H(0,-,0;i)&=&-\frac{i \pi ^3}{48}+\frac{3 \zeta (3)}{8}\;,\\
 H(0,0,+;i)&=&\frac{i}{32} \left(\zeta \left(3,\frac{1}{4}\right)-\zeta \left(3,\frac{3}{4}\right)\right)\;,\\
 H(0,0,-;i)&=&-\frac{3 \zeta (3)}{16}\;,\\
 H(0,0,0;i)&=&-\frac{i \pi ^3}{48}\;.
\end{eqnarray}
\subsection{Arbitrary weight}
\begin{eqnarray}
H({}^n+;i)&=&\frac{(i\pi)^n}{2^n n!}\;,\\
H({}^n-;i)&=&\frac{(-2\log2)^n}{n!}\;,\\
H({}^n0;i)&=&\frac{(i\pi)^n}{2^n n!}\;,\\
H({}^n 0,+;i)&=&i2^{2n+1}\left(\zeta\left(n+1,\frac{1}{4}\right)-\zeta\left(n+1,\frac{3}{4}\right)\right)\;.
\end{eqnarray}
The HPLs at $i$ with only weights $-$ and $0$ can be converted to HPLs at $x=-1$ using the transformation $x\rightarrow x^2$. For weight vectors $s_1,...,s_n$ we have 
\begin{eqnarray}
H(s_1,...,s_n;i)&=&H(s'_1,...,s'_n;-1)
\qquad s_n=-, \quad s_i=0,- s_i'=\left\{\begin{array}{cc}1,&s_i=-\\0,&s_i=0\;,\end{array}\right.\nonumber\\
\end{eqnarray}
so long as the last element of the weight vector is not a 0. If it is so, one has to extract the $0$s from the weight vector using the product identities.
\section{Multiple Zeta Values}\label{sec:MZVtable}
 We list here some of the identities for MZVs found in \cite{math.CA/9910045,Borwein:1996yq}.
\begin{eqnarray}
\zeta(2,1)&=&\zeta(3)\;,\\
\zeta(4,2)&=&\zeta^2(3)-\frac{4\pi^6}{2835}\;,\\
2 \zeta(m,1)&=&m\zeta(m+1)-\sum\limits_{k=1}^{m-2}\zeta(m-k) \zeta(k+1),\;\; 2\le m \in \mathds{Z}\;,\\
\zeta(2,{}^n1)&=&\zeta(n+2)\;,\\
\zeta(3,{}^n1)&=&\zeta(n+2,1)\nonumber\\
&=&\frac{n+2}{2}\zeta(n+3)-\frac{1}{2}\sum_{k=1}^n\zeta(k+1)\zeta(n+2-k)\;,\\
\zeta({}^n2)&=&\frac{2 (2 \pi)^{2n}}{(2n+1)!}\left(\frac{1}{2}\right)^{2n+1}\;,\\
\zeta({}^n4)&=&\frac{4 (2 \pi)^{4n}}{(4n+2)!}\left(\frac{1}{2}\right)^{2n+1}\;,\\
\zeta({}^n6)&=&\frac{6 (2 \pi)^{6n}}{(6n+3)!}\;,\\
\zeta({}^n8)&=&\frac{8 (2 \pi)^{8n}}{(8n+4)!}\left\{\left(1+\frac{1}{\sqrt{2}}\right)^{4n+2}+\left(1-\frac{1}{\sqrt{2}}\right)^{4n+2}\right\}\;,\\
\zeta({}^n\{3,1\})&=&4^{-n}\zeta({}^n4)=\frac{2\pi^{4n}}{(4n+2)!}\;,\\
\zeta(2,{}^n\{1,3\}) &=&4^{-n}\sum_{k=0}^n(-1)^k\zeta(\{4\}^n)\big\{(4k+1)\zeta(4k+2)\nonumber\\
&&\quad\quad-4\sum_{j=1}^k\zeta(4j-1)\zeta(4k-4j+3)\big\}\;,
\end{eqnarray}
\subsection{Colored MZVs}
Here we list some identities for MZV's with negative arguments found in \cite{Borwein:1996yq},
\begin{eqnarray}
\zeta({}^n(-2))&=&\frac{\pi^{2n}}{(2n+1)!}\frac{(-1)^{n(n+1)/2}}{2^n}\;,\\
\zeta({}^n(-4))&=&\frac{\pi^{4n}}{(4n+2)!}\frac{(-1)^{n(n+1)/2}}{2^n}\left(\left(1+\sqrt{2}\right)^{2n+1}+\left(1-\sqrt{2}\right)^{2n+1}\right)\;,\nonumber\\\\
\zeta({}^n(-6))&=&\frac{\pi^{6n}}{(6n+3)!}\frac{3}{2}\Big(1+2^{3n+1}(-1)^{n(n+1)/2}\;\nonumber\\
&&\times\left\{\left(\frac{1+\sqrt{3}}{2}\right)^{6n+3}+\left(\frac{1-\sqrt{3}}{2}\right)^{6n+3}-1\right\}\;,\\
\zeta(-1,{}^n1)&=&(-1)^{n+1}\frac{(\log2)^n}{n!}\;,\nonumber\\
\zeta(-1,-1,{}^n1)&=&-\mathrm{Li}_{n+2}(1/2)\;,\\
\zeta({}^n(-1))&=&(-1)^n\sum\prod_{k\ge 1}\frac{1}{j_k!}\left(\frac{-\mathrm{Li}_k((-1)^k)}{k}\right)^{j_k}\;.
\end{eqnarray}
where the sum in the last equation is over all non negative integers satisfying $\sum_{k\ge 0}k j_k=n$
\subsection{Minimal set}\label{MZVindep}
MZV's can be expressed as linear combination of mathematical constants ($\pi,\zeta(n),\mathrm{Li}_n(1/2)$) and, for high weight, a minimal set of other constants. In the tables we translated\cite{Remiddi}, the following choice of constants in the minimal set has been used.  
\begin{eqnarray}
s_6  &=& S(\{-5,-1\},\infty) =\sum\limits_{i_1=1}^{\infty}\frac{(-1)^{i_1}}{i_1^5} \sum\limits_{i_2=1}^{i_1}\frac{(-1)^{i_2}}{i_2}\nonumber\\
&=&\zeta(-5,-1)+\zeta(6)\simeq 0.98744142640329971377\;,\\
s_{7a}  &=& S(\{-5,1,1\},\infty) =\sum\limits_{i_1=1}^{\infty}\frac{(-1)^{i_1}}{i_1^5} \sum\limits_{i_2=1}^{i_1}\frac{1}{i_2}\sum\limits_{i_3=1}^{i_2}\frac{1}{i_3}\nonumber\\
&=&\zeta(-5,1,1)+\zeta(-6,1)+\zeta(-5,2)+\zeta(-7)\\
&\simeq& -0.95296007575629860341\;,\nonumber\\
s_{7b}  &=& S(\{5,-1,-1\},\infty) =\sum\limits_{i_1=1}^{\infty}\frac{1}{i_1^5} \sum\limits_{i_2=1}^{i_1}\frac{(-1)^{i_2}}{i_2}\sum\limits_{i_3=1}^{i_2}\frac{(-1)^{i_3}}{i_3}\nonumber\\
&=&\zeta(7)+\zeta(5,2)+\zeta(-6,-1)+\zeta(5,-1,-1)\nonumber\\
&\simeq&  1.02912126296432453422\;,
\end{eqnarray}
\begin{eqnarray}
s_{8a}  &=& S(\{5,3\},\infty) =\sum\limits_{i_1=1}^{\infty}\frac{1}{i_1^5} \sum\limits_{i_2=1}^{i_1}\frac{1}{i_2^3}\nonumber\\
&=&\zeta(8)+\zeta(5,3)\simeq 1.0417850291827918834\;,\nonumber\\
s_{8b}  &=& S(\{-7,-1\},\infty) =\sum\limits_{i_1=1}^{\infty}\frac{(-1)^{i_1}}{i_1^7} \sum\limits_{i_2=1}^{i_1}\frac{(-1)^{i_2}}{i_2}\nonumber\\
&=&\zeta(8)+\zeta(-7,-1)\simeq 0.99644774839783766600\;,\\
s_{8c}  &=& S(\{-5,-1,-1,-1\},\infty) \nonumber\\&=&\sum\limits_{i_1=1}^{\infty}\frac{(-1)^{i_1}}{i_1^5} \sum\limits_{i_2=1}^{i_1}\frac{(-1)^{i_2}}{i_2}\sum\limits_{i_3=1}^{i_2}\frac{(-1)^{i_3}}{i_3}\sum\limits_{i_3=1}^{i_3}\frac{(-1)^{i_4}}{i_4}\nonumber\\
&=&\zeta(8)+\zeta(-7,-1)+\zeta(-5,-3)+\zeta(6,2)+\zeta(-5,-1,2)\nonumber\\
&&+\zeta(-5,2,-1)+\zeta(6,-1,-1)+\zeta(-5,-1,-1,-1)\\
&\simeq& 0.98396667382173367094\;,\nonumber\\
s_{8d}  &=& S(\{-5,-1,1,1\},\infty) =\sum\limits_{i_1=1}^{\infty}\frac{(-1)^{i_1}}{i_1^5} \sum\limits_{i_2=1}^{i_1}\frac{(-1)^{i_2}}{i_2}\sum\limits_{i_3=1}^{i_2}\frac{1}{i_3}\sum\limits_{i_3=1}^{i_3}\frac{i}{i_4}\nonumber\\
&=&\zeta(8)+\zeta(-5,-3)+\zeta(6,2)+\zeta(7,1)+\zeta(-5,-2,1)+\zeta(-5,-1,2)\nonumber\\
&&+\zeta(6,1,1)+\zeta(-5,-1,1,1)\simeq 0.99996261346268344768\;.
\end{eqnarray}
\section{Table of representation through more common functions}\label{sec:identities}
In these tables, we collected the identities relating HPLs to more common functions that can be found in \cite{Moch:1999eb,Gehrmann:2000zt}. These identities are used by the function {\tt HPLConvertToKnownFunctions} (see section \ref{sec:newfunctions})and applied systematically when the option {\tt \$HPLAutoConvertToKnownFunctions} is set {\tt True}, see section \ref{sec:options}. 
\subsection{Weight 2}
\begin{eqnarray}
H\left(\{1, 1\}; x\right) &=& \frac{1}{2} \log\left(1 - x\right)^2 \;,\\
H\left(\{1,-1\};x\right)&=&
  {\mathrm{Li}}_{2}((1-x)/2)-\log(2) \log(1-x)-\mathrm{Li}_{2}\left(\frac{1}{2}\right)\;,\\
H\left(\{-1,1\};x\right)&=&
  \mathrm{Li}_{2}((1+x)/2)-\log(2) \log(1+x)-
      \mathrm{Li}_{2}\left(\frac{1}{2}\right)\;,\\
H\left(\{-1,0\};x\right)&=&
  \log(1+x) \log(x)+\mathrm{Li}_{2}(-x)\;,\\
H\left(\{-2\};x\right)&=&-\mathrm{Li}_{2}(-x)\;.
\end{eqnarray}
\subsection{Weight 3}
\begin{eqnarray}
H\left(\{1, 2\}; x\right) &=& -2 S_{1,2}\left( x\right) -\log\left(1 - x\right) \mathrm{Li}_2\left( x\right) \;, \\
H\left(\{1, 1, 1\};x\right) &=& -\frac{1}{6} \log\left(1 - x\right)^3 \;, \\
H\left(\{2, 1\}; x\right) &=& S_{1,2}\left( x\right)  \;,\\
H\left(\{2,-1\};x\right)&=&  \mathrm{Li}_{3}(\frac{2x}{1+x})-\mathrm{Li}_{3}\left(\frac{x}{1+x}\right)-\mathrm{Li}_{3}\left(\frac{1+x}{2}\right)\nonumber\\
&&-\mathrm{Li}_{3}(x)+\log(1+x) \mathrm{Li}_{2}(\frac{1}{2})+\log(1+x) \mathrm{Li}_{2}(x)\nonumber\\
&&+\frac{1}{2}  \log(2) \log(1+x)^2+\mathrm{Li}_{3}\left(\frac{1}{2}\right)\;,\\
H\left(\{-2,1\};x\right)&=&-S_{1,2}(x)+\mathrm{Li}_{3}\left(\frac{-2 x}{1-x}\right)-\mathrm{Li}_{3}\left(\frac{1-x}{2}\right)-\mathrm{Li}_{3}(-x)\nonumber\\
&&+\mathrm{Li}_{3}\left(\frac{1}{2}\right)+\mathrm{Li}_{3}(x)+\log(1-x) \mathrm{Li}_{2}(-x)\nonumber\\
&&+\log(1-x) \mathrm{Li}_{2}\left(\frac{1}{2}\right)-\log(1-x) \mathrm{Li}_{2}(x)\nonumber\\
&&+\frac{1}{2} \log(2) \log(1-x)^2-1/6 \log(1-x)^3\;,\\
H\left(\{-2,-1\};x\right)&=&S_{1,2}(-x)\;,\\
H\left(\{1,-1,-1\};x\right)&=&-\frac{1}{2} \log\left(\frac{1-x}{2}\right) \log(1+x)^2-\mathrm{Li}_{3}\left(\frac{1}{2}\right)\nonumber\\
&&-\log(1+x) \mathrm{Li}_{2}\left(\frac{1+x}{2}\right)+\mathrm{Li}_{3}\left(\frac{1+x}{2}\right)\;.
\end{eqnarray}
\begin{eqnarray}
H\left(\{-1,1,1\};x\right)&=&  \frac{1}{2} \log\left(\frac{1+x}{2}\right) \log(1-x)^2+\mathrm{Li}_{3}\left(\frac{1}{2}\right)\nonumber\\
&&+\log(1-x) \mathrm{Li}_{2}\left(\frac{1-x}{2}\right)-      \mathrm{Li}_{3}\left(\frac{1-x}{2}\right)\;,\\
H\left(\{-1,-2\};x\right)&=&  H\left(\{-2\};x\right) H\left(\{-1\};x\right)-2  H\left(\{-2,-1\};x\right)\;,\\
H\left(\{-1,2\};x\right)&=&  H\left(\{-1\};x\right) H\left(\{2\};x\right)-H\left(\{-2,1\};x\right)\nonumber\\
&&-H\left(\{2,-1\};x\right)\;,\\
H\left(\{-1,1,-1\};x\right)&=&  H\left(\{-1\};x\right) H\left(\{1,-1\};x\right)-2  H\left(\{1,-1,-1\};x\right)\;.\nonumber\\ 
\end{eqnarray}
\subsection{Weight 4}
\begin{eqnarray}
H\left(\{1, 3\}; 
    x\right) &=& -\frac{1}{2}  \mathrm{Li}_2\left(x\right)^2 - 
      \log\left(1 - x\right) \mathrm{Li}_3\left( x\right) \;, \\
H\left(\{1, 1, 2\}; x\right) &=& 
  \frac{1}{2} \log\left(1 - x\right)^2 \mathrm{Li}_2\left( x\right) + 2 \log\left(1 - x\right) S_{1,2}\left( x\right) \nonumber\\
  &&+ 
      3 S_{1,3}\left( x\right) \;, \\
H\left(\{1, 2, 1\}; 
    x\right) &=& -\log\left(1 - x\right) S_{1,2}\left( x\right) - 
      3 S_{1,3}\left( x\right)\;,  \\
H\left(\{2, 2\}; 
    x\right) &=& -2 S_{2,2}\left( x\right) + 
      \frac{1}{2} \mathrm{Li}_2\left(x\right)^2 \;,\\  
H\left(\{3, 1\}; x\right) &=& S_{2,2}\left( x\right)\;,\\  
H\left(\{2, 1, 1\}; x\right) &=& 
  S_{1,3}\left( x\right)\;,\\  
H\left(\{1, 1, 1, 1\}; x\right) &=& 
  \frac{1}{24} \log\left(1 - x\right)^4  \;.
\end{eqnarray}
\subsection{Arbitrary weight}
\begin{eqnarray}
H\left(\{n\}; x\right) &=& 
  \mathrm{Li}_n\left(x\right),\qquad  n>0\;,\\
 H\left(\{n\}; x\right) &=& 
  -\mathrm{Li}_{-n}\left(-x\right),\qquad  n<0\;,\\   
H(\{^n 1\}; 
    x) &=& (-1)^n \frac{\log\left(1 - x\right)^n}{n!} \;,\\
H(\{^n(-1)\}; 
    x) &=& \frac{\log\left(1+x\right)^n}{n!} \;, \\
H(\{n,^{p-1}1\}; x) &=& 
  S_{n-1,p}\left( x\right)\;.  
\end{eqnarray}








\end{document}